\documentclass[number,citesort,dvips]{arxstspdf}
\usepackage{graphics,flushend}
\volume{25}
\issue{1}
\pubyear{2010}
\firstpage{126}
\lastpage{144}
\doi{10.1214/09-STS283}

\makeatletter
\setattribute{abstract}    {skip} {20\p@}
\setattribute{keyword}     {skip} {1\p@}
\setattribute{frontmatter} {skip} {0\p@ plus 3\p@ minus 3\p@}
\setattribute{frontmatter} {cmd}  {}
\setattribute{abstract}   {width}  {41.5pc}
\setattribute{keyword}    {width}  {41.5pc}

\makeatother

\begin{document}
\begin{frontmatter}

\title{A Conversation with James Hannan}
\runtitle{A Conversation with James Hannan}

\begin{aug}
\author[a]{\fnms{Dennis} \snm{Gilliland}\ead[label=e1,text=gilliland@ stt.msu.edu]{gilliland@stt.msu.edu}\corref{}}
\and
\author[b]{\fnms{R. V.} \snm{Ramamoorthi}\ead[label=e2]{ramamoor@stt.msu.edu}}
\runauthor{D. Gilliland and R. V. Ramamoorthi}

\affiliation{Michigan State University}

\address[a]{Dennis Gilliland is Professor, Department of Statistics
and Probability,
Michigan State University \printead{e1}.}
\address[b]{R. V. Ramamoorthi is Professor, Department of Statistics
and Probability,
Michigan State University \printead{e2}.}

\freefootnotetext[]{This conversation was conducted in spring 2008.
We are sad to report that Jim passed away on January 26, 2010.
An obituary appears in the March 2010 issue of the IMS Bulletin.}
\end{aug}

%

\begin{abstract}\fontsize{8.6}{9.6}\selectfont{
Jim Hannan is a professor who has lived an interesting
life and one whose fundamental research in repeated games
was not fully appreciated until late in his career. During
his service as a meteorologist in the Army in World War II,
Jim played poker and made weather forecasts. It is curious that
his later research included strategies for repeated play that apply
to selecting the best forecaster.

James Hannan was born in Holyoke, Massachusetts on September 14, 1922.
He attended St. Jerome's High School and in January 1943 received the
Ph.B. from St. Michael's College in Colchester, Vermont. Jim enlisted
in the US Army Air Force to train and serve as a meteorologist. This
took him to army airbases in China by the close of the war. Following
discharge from the army, Jim studied mathematics at Harvard and graduated
with the M.S. in June 1947. To prepare for doctoral work in statistics
at the University of North Carolina that fall, Jim went to the University
of Michigan in the summer of 1947. The routine admissions' physical revealed
a spot on the lung and the possibility of tuberculosis. This caused
Jim to stay at Ann Arbor through the fall of 1947 and then at a Veterans
Administration Hospital in Framingham, Massachusetts to have his condition
followed more closely. He was discharged from the hospital in the spring
and started his study at Chapel Hill in the fall of 1948. There he began
research in compound decision theory under Herbert Robbins. Feeling the
need for teaching experience, Jim left Chapel Hill after two years and
short of thesis to take a three year appointment as an instructor at
Catholic University in Washington, DC. When told that renewal was not
coming, Jim felt pressure to finish his degree. His 1953 UNC
thesis contains results in compound decision theory,
a density central limit theorem for the generalized binomial and exact
and asymptotic distributions associated with a Kolmogorov statistic.
He was encouraged to apply to the Department of Mathematics at Michigan
State University and came as assistant professor in the fall of 1953.
In the next few years, he accomplished his work on repeated games. The
significance of the work was rediscovered by the on-line learning
communities in computer science in the 1990s and the term \textit
{Hannan consistency}
was coined. His retirement came in 2002 after a long career that
included major contributions to compound and empirical Bayes decision
theory and other areas. He and his colleague V\'{a}clav Fabian
co-authored \textit{Introduction to Probability and Mathematical Statistics}
(Wiley 1985).

A \textit{Hannan strategy} is a strategy for the repeated play of a game
that at each stage $i$ plays a smoothed version of a component Bayes rule
versus the empirical distribution Gi-1 of opponent's past plays. [Play
against the unsmoothed version is often called (one-sided) \textit
{fictitious play}.]
As in compound decision theory, performance is measured in terms
of \textit{modified regret}, that is, excess of average risk across
stages $i = 1,\ldots,n$ over the component game Bayes envelope R evaluated
at Gn. Hannan, James F., Approximation to Bayes Risk in
Repeated Play, \textit{Contributions to the Theory of Games} \textbf
{3} 97--139,
Princeton University Press, is a paper rich with bounds on modified
regrets. A \textit{Hannan consistent} strategy is one where
limsup (modified regret) is not greater than zero. In the 1990s,
greater recognition of Hannan's work began to emerge; the term Hannan
consistency may have first appeared in Hart and Mas-Colell
[\textit{J. Econom. Theory} \textbf{98} (2001) 26--54].

Early on, only his students and a few others were aware of the
specifics of his findings. The failure of others to recognize the specific
results in the 1957 paper may be due to the cryptic writing style and notation
of the author. The strategy for selecting forecasters in Foster and
Vohra \cite{foster} [\textit{Operations Research} \textbf{41} (1993) 704--709]
is an unrecognized Hannan-strategy as is the strategy
in Feder et al. \cite{feder} [\textit{IEEE Trans. Inform. Theory} \textbf{38} (1992) 1258--1270].
Gina Kolata's \textit{New York Times} article, ``Pity the Scientist
who Discovers the Discovered'' (February 5, 2006) uses the original Hannan
discoveries as an example, although referring to him as a
``statistician named James
Hanna.''

In May 1998, the Department of Statistics and Probability hosted a
\textit{Research Meeting in Mathematical Statistics in Honor of
Professor James Hannan}.
Many came to honor Jim; the speakers included V\'{a}clav Fabian,
Stephen Vardeman,
Suman Majumdar, Richard Dudley, Yoav Freund, Dean Foster, Rafail Khasminskii,
Herman Chernoff, Michael Woodroofe, Somnath Datta, Anton Schick and
Valentin Petrov.

Jim was ever generous in giving help to students.
He enjoyed improving results and was very reluctant to submit research results
until much effort was made to improve them. Jim directed or co-directed the
doctoral research of twenty students: William Harkness (1958),
Shashikala Sukatme (1960), John Van Ryzin (1964), Dennis Gilliland (1966),
David Macky (1966), Richard Fox (1968), Allen Oaten (1969), Jin Huang (1970),
Vyagherswarudu Susarla (1970), Benito Yu (1971), Radhey Singh (1974),
Yoshiko Nogami (1975), Stephen Vardeman (1975), Somnath Datta (1988),
Jagadish Gogate (1989), Chitra Gunawardena (1989), Mostafa Mashayekhi (1990),
Suman Majumdar (1992), Jin Zhu (1992) and Zhihui Liu (1997). Most
pursued academic
careers and some ended up at research universities including
Pennsylvania State,
Columbia, Michigan State, UC-Santa Barbara, Guelph, SUNY-Binghamton,
Iowa State,
Louisville, Nebraska-Lincoln and Connecticut-Stamford.

It was in the Army in 1944 that Jim read his first statistics book.
It was \textit{War Department Education Manual EM 327, An Introduction to
Statistical Analysis}, by C. H. Richardson, Professor of Mathematics,
Bucknell University, published by United States Armed Forces Institute,
Madison, Wisconsin (CQ).}
\end{abstract}

%

\end{frontmatter}

\section{St. Jerome High School 1935--1939}
\textbf{Q.} Where were you raised?

\textbf{A.} I was raised in Holyoke, Massachusetts by my father and my mother's
sister. I was an only child; my mother died shortly after my birth. I
went to St. Jerome's High School in Holyoke, a long block from where I
was living. The juniors and seniors preparing for college were combined
into one class for two years of courses. The second year courses could
not build upon the first year courses.

\textbf{Q.} I recall meeting you for the first time at a statistics department
picnic in 1963. I was impressed by the fact that you brought a glove
and the way you played the game. Did you participate in sports in high school?

\textbf{A.} The debate coach (priest) was the baseball coach, and he was
enthralled with the logical way I laid out a debate. He was an
ex-pitcher who started coaching me in pitching. I think he was
impressed with the logic of my arguments. I had a basic knowledge of
Aristotelian logic without realizing it. I was on the debate team and
the baseball team.

\textbf{Q.} We have seen your debating skills exhibited in faculty meetings over
the years. Were you the best student in your high school class?

\textbf{A.} Probably so.

\section{St. Michael'S College Sept 1939--Dec 1942}
\textbf{Q.} Did you go to Mt. Holyoke College?

\textbf{A.} No. The only time I had anything to do with Mt. Holyoke College was
when I went there to take the Graduate Record Examination or something
like that.

\begin{figure}

\includegraphics{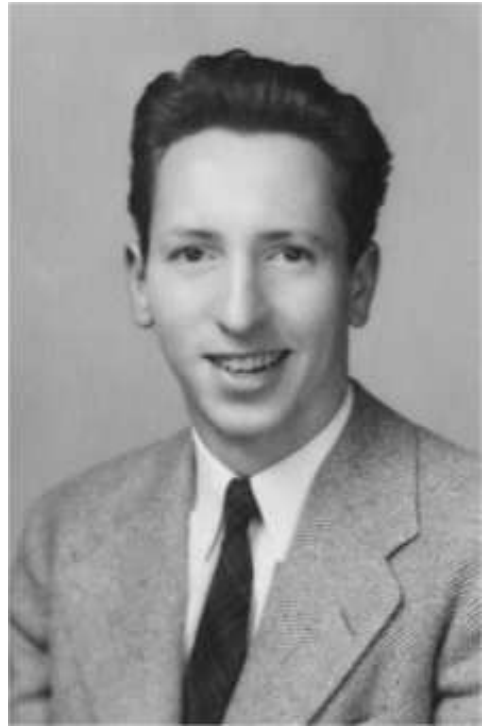}

  \caption{Jim at St. Michael's College, 1942.}
\end{figure}

\textbf{Q.} Where did you go to college following high school graduation in 1939?

\textbf{A.} I went to a small college in Vermont. I had a scholarship from my
high school that was only payable at a catholic college. I went to St.
Michael's College, which is about a mile from Winooski which is about
two miles from Burlington (laugh).

\textbf{Q.} Did you spend 4 years at St. Michael's?

\textbf{A.} Well, when the United States entered the war in December 1941,
colleges introduced paths for early graduation. Summer programs and
accelerated\break courses were introduced. By the end of the fall of 1942, I
had earned 25\% more than the required graduation credits and did not
bother to attend the last month of class. My degree came formally in
January 1943.

\textbf{Q.} What was the degree and major?

\textbf{A.} I earned the bachelor of philosophy (Ph.B.) degree and was a
mathematics major. I was a pathetic math major (laugh). I avoided
extensive coursework in classical languages and the study of church
religion in Latin by taking the Ph.B. degree rather than the B.A. degree.

\textbf{Q.} What was your exposure to mathematics at St. Michael's?

\textbf{A.} There was a first class mathematics instructor at St. Michael's when
I first went there in 1939, but he disappeared the next year. His name
was Andr\'{e} Gleyzal. He taught me pre-calculus from Sutherland
Frame's book. Gleyzal was not interested in doing any extra work. He
simply had his grader stamp your homework AC, meaning accepted.

\textbf{Q.} Was there a teaching style that influenced you? Were the mathematics
courses rigorous at St. Mi\-chael's?

\textbf{A.} There was an attempt at rigor, probably in differential equations in
the third year. And the man who was teaching it didn't understand it (laugh).

\textbf{Q.} This was clear to you and few others?

\textbf{A.} This was pretty clear. He was from St. Louis University and working
in some area of geometry and this is not much of a recommendation for
his analysis.

\textbf{Q.} Were your extra credits at St. Michael's in mathematics?

\textbf{A.} There were not many math courses there. My last year I was taking
courses in solid and analytic geometry and some course in classical
algebra. They didn't have a regular faculty member in mathematics then;
there was a graduate of St. Michael's who served as part registrar and
math teacher. He was a bright guy, but he had no perspective.

\textbf{Q.} Had you heard of statistics by the time you graduated from college?

\textbf{A.} No.

\textbf{Q.} How many students were at St. Michael's then?

\textbf{A.} I would say about 125 students, now it has twenty times that number
I suppose.

\textbf{Q.} Did you go back and forth to home very often?

\textbf{A.} It was a long afternoon hitching rides. With no success by evening,
we headed to the train depot.

\textbf{Q.} You mentioned that the US entry into the war accelerated your
undergraduate education. How did this work?

\textbf{A.} St. Michael's offered fall semester courses in the summer and used
the fall for courses that would ordinarily be offered in the spring.

\textbf{Q.} Was the intervention to get students through so that they could serve?

\textbf{A.} It was probably more the case that they were helping students
graduate before their deferments expired.

\section{US Army Dec 1942--Dec 1945}

\textbf{Q.} Did you go directly from college into the Army?

\begin{figure}

\includegraphics{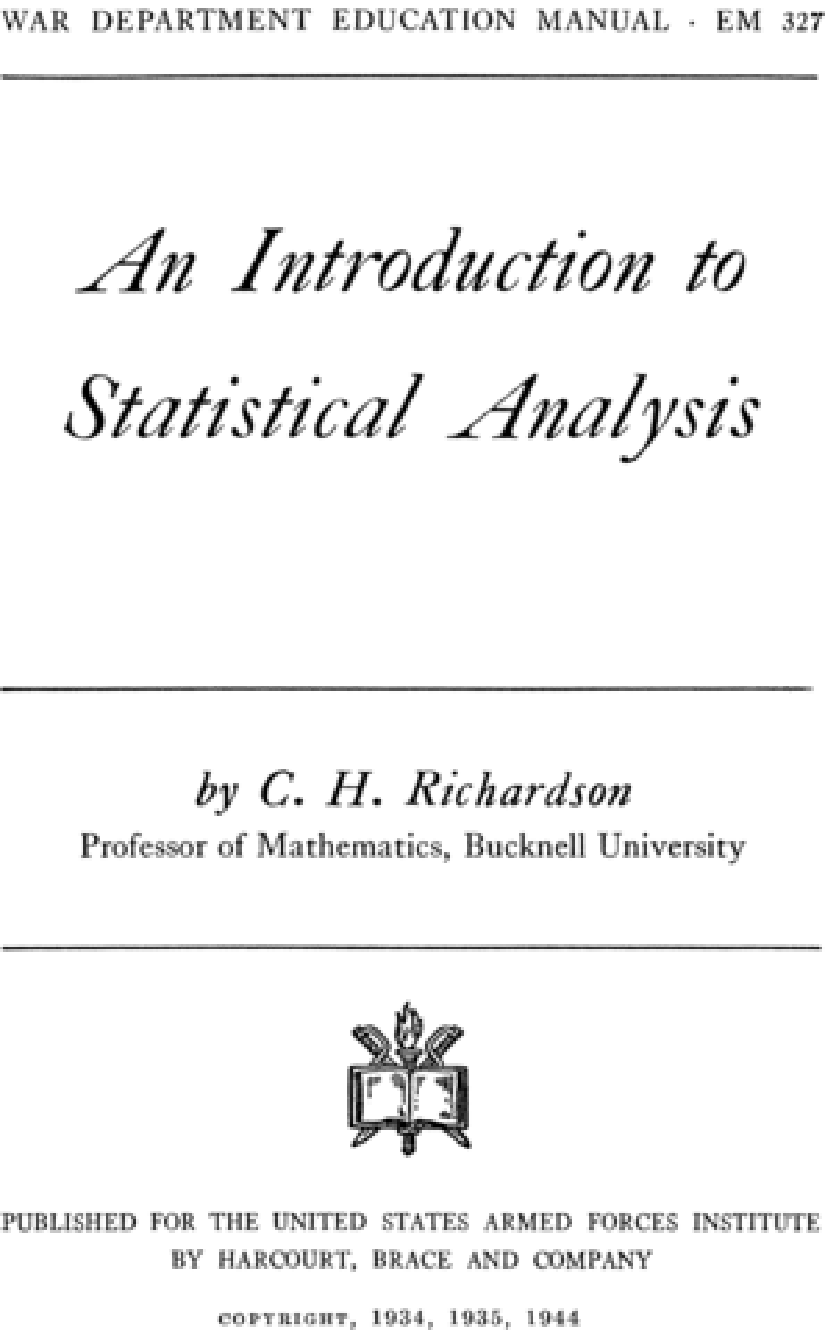}

  \caption{Jim's Army Manual in Statistics, 1944.}
\end{figure}

\textbf{A.} As I mentioned earlier, I stopped attending classes in December 1942.
I managed to bypass my draft board by volunteering for training in meteorology
in the Army. The training program in meteorology was nine months, the longest
one that the Army offered (laugh). I was told that the training was to
take place
at one of four prestigious universities, either Massachusetts Institute
of Technology,
California Institute of Technology, The University of Chicago, I have
forgotten the fourth.

\textbf{Q.} Where did you go for training?

\textbf{A.} It was not at any of the aforementioned universities. It was at the
civic center in Grand Rapids, Michigan. No explanation was given to us.
It was clear that the Army had made some mistake in training numbers
and was forced to send the recruits to places where they received less
training. The civic center had been taken over by the Army, as was a
downtown hotel that served as the dormitory.

\textbf{Q.} Did it go the full nine months as promised?

\textbf{A.} Yes, but study was not very compatible with the required physical
training. We were so exhausted after physical training that we went to
our rooms to sleep; there was no study.
Classes would end at 4 pm. Then they would march us in our units
through the Grand Rapids streets up to John Ball Park for drill. I
qualified with the carbine, both in shooting and in stripping the
weapon. I also qualified with the 45 caliber hand gun. You couldn't hit
a barn with it, but got used to the recoil.

\textbf{Q.} How many were in your cohort?

\textbf{A.} I think that there were about 300 of us. The Army had a handle on
mass production. The Army gave us examinations and I think that I
turned out to have done fairly well in mathematics, surprisingly.

\textbf{Q.} Did you rank first on the mathematics exam?

\textbf{A.} No, I didn't but I scored in the top 10\%. This was a surprise to
me. Among the trainees were men from the East Coast and New York City
who were much better educated in mathematics than I was. Apparently I
created more of a good impression with the Army by scoring rather high
on an IQ test.

\textbf{Q.} When were you commissioned as a Second Lieutenant?

\textbf{A.} Upon completion of training in the fall of 1943. The Army found out
by this time that it had trained more meteorologists than it needed.

\textbf{Q.} We presume that this could have been bad news.

\textbf{A.} Yes, but the Sergeant had already turned in the roster with me
assigned to the Army Airbase at Lewiston, Montana. When I arrived at
Lewiston, I found that the base was five miles away. I had never
learned to drive. However, people put me in a Jeep, pointed me in the
right direction, turned on the key, and told me to turn off the key
when I arrived at the base.

\textbf{Q.} What was your position at the base?

\textbf{A.} I was already commissioned and qualified as a meteorologist. At the
base, I was supervised by the person who ran the weather station. The
work as a meteorologist was made somewhat complicated by the fact that
the information and forecasts had to be coded. The Army thought that
someone could be listening and learning something about the weather.
There was a different code for every day.

\textbf{Q.} We suppose that you created graphs with symbols indicating various
information, for example, arrows indicating wind velocity.

\textbf{A.} Yes, I did. We used the Beaufort system using number of dashes for
wind speed. I had learned the system but not with ease.

\textbf{Q.} What was life like on the base?

\textbf{A.} We lived on the base in wooden barracks. There was an officers' mess
hall. There a second lieutenant does not get much respect, none from
below, little from above.

\textbf{Q.} How long were you at Lewistown?

\textbf{A.} Only about four months. For some reason, I applied for tropical
weather training. The Army pulled me out of Lewistown about Christmas
and sent me to Miami, Florida to await duty assignment in San Juan,
Puerto Rico. I then spent three or four months there in fulltime
advanced study of tropical meteorology.

\textbf{Q.} Did you ever get feedback on your forecasts based on the outcomes?

\textbf{A.} (Laugh) Yes, I recall one case when I was back in the United States.
I gave a notoriously bad forecast. I had forecasted that the low cloud
cover would rise up and blow away. The pilot took off based on my
forecast and radioed back from ten thousand feet that it was solid all
the way up.

\textbf{Q.} Was the feedback from outcomes useful?

\textbf{A.} Yes, but mainly it helped us recognize how far off forecasts could be.

\textbf{Q.} What was your next assignment?

\textbf{A.} I was sent by ship to Bombay (Mumbai), India. I recall being in the
harbor there in May 1944 after a very long ocean voyage. Along the way
we stopped in Algeria and waited a week to transfer to a British ship
for the remainder of the journey. As we traversed the Mediterranean, I
served as officer of the day once. This was not duty that one sought.
You were required to inspect the prisoners in the brig and every nook
and cranny on the ship and be on deck while everyone else was posted below.

\textbf{Q.} What happened following your leave from the ship in the harbor at Bombay?

\textbf{A.} Rooms were provided in the city. Orders arrived in a week starting
with a three day trip across India to Calcutta (Kolkata) on a troop
train. The train was rather open to the elements. It was very hot and
at the onset of the monsoon season. Rain blew into the train. I was as
sick as a dog; practically everyone had an infection of sorts. When we
arrived, I did not realize that I was in the vicinity of a place of
statistical significance, the Mahalanobis Estate in Barrackpore. The
Estate was eventually turned-over to the Indian Statistical Institute.

\textbf{Q.} Have you ever visited the Indian Statistical Institute in Barrackpore?

\textbf{A.} No.

\textbf{Q.} How long did you stay in Calcutta?

\textbf{A.} I probably stayed a week or two in Calcutta before being sent to a
small village, Gushkara, that was 150--200 miles to the northwest.
There I was in charge of the meteorology set-up. I was the only officer
at the weather station. I only spent a few months in the village. The
village had an air strip and served flights in and out of China. It was
a pretty quiet base. Information came in over the radio in code.
Teletype had not reached the village yet.

\textbf{Q.} Do you have other recollections of your stay at Gushkara?

\textbf{A.} I recall that I had a dentist friend there on the base who fixed up
my teeth with early morning appointments.

\textbf{Q.} At some cost?

\textbf{A.} No, because he was a friend and a fellow bridge player. Actually, it
was poker that served me well in the army. My poker playing started on
the trip to Bombay. There were poker games below deck. There were games
going the entire journey. I was forced to leave the games that one day
when I was officer of the day while traveling across the Mediterranean.
It was pretty much just playing poker, having regular meals and sleep.
I would stay below deck for days at a time. Some of the games had pots
that built up to a couple of hundred dollars.

\textbf{Q.} Apparently, you did not get to see much scenery.

\textbf{A.} Well, going through the Suez Canal I did once venture on deck. About
all one could see from the ship were the banks of the Canal.

\textbf{Q.} Did the Army discourage this gambling?

\textbf{A.} No. In fact, at every base I was at there were card games going in
the officers' mess. You could pretty much find a game whenever you wanted.

\textbf{Q.} Where did your next orders take you?

\textbf{A.} I was ordered to proceed to China to join the combined Chinese
American forces. We flew ``over the hump'' to the base at Yunnanyi. I
have kept a small diary that indicates we arrived on March 23. I then
drove a Jeep to the air base at Kunming, arriving on April 12. This was
a major base and far enough from the enemy forces to not be in danger
of being over run. I kept the Jeep to eventually deliver to a base in
Chihkiang (Zhejiang) close to the front lines. I recall that the last
base was within 50 miles of Japanese forces. It was a dangerous place
with enemy forces dressing as coolies to infiltrate the area. From this
base, the air missions were very short and in support of ground
operations. That base had an event that made an issue of Time magazine.
The commanding officer had mercifully shot an American pilot who was
trapped in a burning plane. There was an official court martial but the
commanding officer was exonerated.

\textbf{Q.} What was driving like across China?

\textbf{A.} There were mountain roads with lots of switchbacks. It was a
dangerous trip. It felt as if we were going vertical on the turns and
always coming close to turning over. There were no roll bars on the vehicle.

\textbf{Q.} What were your major responsibilities as a meteorologist at these bases?

\textbf{A.} Generally there were two meteorologists to man the weather station,
each working a 12 hour shift. I sometimes pulled the night shift. Our
responsibilities included analyzing data and giving advisories to
flight crews leaving the base.

\textbf{Q.} Did you ever forecast using a probability, say, 70\% chance of rain?

\textbf{A.} No. But sometimes forecasts were hedged by putting conditions on
them such as ``if this, then this.'' At some point our forecasts came
from searching historical records to find days where conditions matched
those of the current day. We then forecast what was typically observed
the following day.

\textbf{Q.} Something like \textit{play against the past}?

\textbf{A.} Yes. Actually, I believe that the success of weather forecasting
today comes from the availability of computers to solve large systems
of thermodynamic equations. It is a long way from looking at the sky
and putting up your finger.

\textbf{Q.} Were you kept busy?

\textbf{A.} Sometimes we were behind in our work. Other times it was make work
to keep us busy.

\textbf{Q.} When did you depart from China?

\textbf{A.} After about three months at my last base, we received word that the
war had ended. This was in August 1945. They moved us to the coast to
await transport to America. While waiting, I saw Japanese soldiers as
prisoners being marched through the streets. Eventually we were taken
to Shanghai and we departed by ship to Seattle, probably in November.
This ship was much faster than those that took us to Bombay. From
Seattle I went by train to Fort Devins, Massachusetts to be discharged.
You might say, ``Around the world in two years.''

\textbf{Q.} Did you play poker on the trip from Shanghai to Seattle?

\textbf{A.} (Laugh) Probably not. I think that I had experienced some bad nights
in China and by this time was not playing poker.

\textbf{Q.} But you still came back with winnings?

\textbf{A.} Oh, yes. I had banked about two thousand dollars which eventually
helped to support me through graduate programs.

\textbf{Q.} You were commissioned as a Second Lieutenant as you graduated from
your meteorology training. Did you leave the Army as a Second Lieutenant?

\textbf{A.} (Laugh) No, I left as a First Lieutenant; not everyone receives many
promotions as you win a war. In fact, a base commander, a Major, ran
into heavy weather and blamed the weather forecaster. This did not help
my chances for promotion. However, I did leave with a warning about
tuberculosis. I do not remember being too concerned, but I thought
others would be concerned, particularly, those about me.

\textbf{Q.} Did you consider a career in meteorology?

\textbf{A.} No. The training was minimal compared to what may be available today.

\textbf{Q.} What occupied your time until you started graduate school?

\textbf{A.} When I returned to Massachusetts in December 1945, I applied for
Harvard, fall 1946. As a somewhat delayed reaction, I decided to start
in the summer of 1946.

\section{Harvard June 1946--June 1947}
\textbf{Q.} When did you become aware of statistics?

\textbf{A.} I guess there was that Wisconsin Army Text, really an old classic of
sorts. At Harvard I took a two semester course in probability and
statistics from Von Mises. He was the instructor.

\textbf{Q.} Were you working on a master's degree?

\textbf{A.} Yes, I was rather rusty at that point. When I went to Harvard in the
summer of 1946, I took a full program. There was an instructor visiting
from Berkeley by the name of Wolfe. He taught a course in differential
equations and it (laugh) was a real course. I was not very good at it,
but luckily there was a bunch of teachers in the same class and they
were worse (laugh). Well, we were using Ince as a source, a classical
book; and we did construct solutions.

\textbf{Q.} Do you recall the other courses that you took at Harvard?

\textbf{A.} I took a course in projective geometry in the summer of 1946 which I
did not understand very well. I had a full program in the fall of 1946
and the spring of 1947. I had complex analysis from Alfors. I took a
two-semester course in modern algebra from George Mackey who was well
known for his research in various fields of abstract mathematics. We
used the textbook by Birkhoff and Mac Lane.

Then I had another course in the mathematics of physics. Birkhoff ended
up teaching the second semes\-ter of the course. The first semester was
taught by Van Vleck, who was a mathematician of sorts, but he was more
in applied math.

\textbf{Q.} Which Birkhoff taught that second semester of mathematical physics?

\textbf{A.} Garrett Birkhoff. Garrett is famous for his work on lattice theory
and for working with Mac Lane. Garrett's father, George Birkhoff, was
more famous for his work in differential equations, but he wasn't
around at that time.

\textbf{Q.} What was the mathematical physics course like?

\textbf{A.} It was classical Fourier series and special functions.

\textbf{Q.} Was Harvard on the semester system?

\textbf{A.} Yes, but in the summer session there were few regular Harvard
faculty teaching courses; most of the teachers were visitors.

\textbf{Q.} Did you take courses at Harvard in the summer of 1947?

\textbf{A.} No, I graduated in the spring of 1947. I just eked through the
masters. The masters was a condemnation then.

\textbf{Q.} Were you thinking about the Ph.D.?

\textbf{A.} I was thinking about it. Some of my fellow students were intending
to go on for the Ph.D. and a few of them made it.

\textbf{Q.} Having completed the masters, did you consider a nonacademic career path?

\textbf{A.} No, I was intent on pursuing the Ph.D. I applied to about 10 or 12
graduate schools to start study in mathematics in the fall of 1947. The
schools included Columbia, Princeton, Yale and the University of North
Carolina. I was accepted by some, rejected by others. I was accepted
into the Ph.D. program in statistics at the University of North Carolina
for the fall of 1947. I believe that Yale and Princeton turned me down.
Later I met persons at Princeton and they explained that Princeton had
a small quota. They could fill the quota with applicants much better
that I was.

\textbf{Q.} Had you decided to concentrate on a particular area of mathematics?

\textbf{A.} No, I had no idea, but I had been influenced by Von Mises at
Harvard. He was a professor of aeronautical engineering. He was
serious, deliberate and insistent upon following his approach. He was
more into the philosophy of randomness and foundations. He had
postulates but the stuff didn't work. His course did lead us slightly
into statistics.
\section{University of Michigan June~1947--Dec~1947}
\textbf{Q.} Why did you go to the University of Michigan in the summer of 1947?

\textbf{A.} I realized that I needed more course work before starting my
doctoral program and went to Ann Arbor in the summer of 1947. It was
one of a few universities that I knew of that offered summer courses of
interest to me. I figured that it would be good preparation. I~went
there with the intention of staying for the summer only, but the
routine, required physical examination revealed the spot on my lung and
the possibility of tuberculosis. This caused me to stay through the
fall of 1947 as my condition was being followed.

\textbf{Q.} What courses did you take in Ann Arbor?

\textbf{A.} I took some basic math courses. One was a second year modern algebra
course and another was a course in group representations by Richard
Brauer, which I dropped later. Probably, I had to enroll in a number of
courses to keep my GI benefits. There were about twenty faculty sitting
in and three students enrolled. The faculty had much more appreciation
of group representations and what they were good for. Brauer was so
smooth that you could leave class convinced that you knew the subject
when you really did not.

\textbf{Q.} You mentioned courses on the algebra side yet your academic career
shows more work on the analysis side.

\textbf{A.} I also took the real variables course from Hildebrand. He was
somewhat of a source of inspiration. I took a statistics course from
Dyer and C. C. Craig. They were both Ann Arbor products. Howard Raiffa
was a student of Dyer\ldots\ Dyer's prize student. I~did not take a
course from probability people.

\textbf{Q.} Was the Dyer and Craig course a mathematical statistics course?

\textbf{A.} It was theoretical from the point of view of nonmathematicians. It
was not very satisfying. C. C. Craig did stimulate me to think about
pooling information. We had a small class. Each day when he came to
class he would rattle coins in his pocket and challenge us to make
estimates concerning the coins in the pocket that day. He was trying to
convince us that one could not use the Bayes theorem in a meaningful
way. (Laugh) It is very hard to prove a negative result.

\textbf{Q.} Any other recollections regarding your stay in Ann Arbor?

\textbf{A.} I played bridge regularly. Once I bid ``three no-trump'' and an
opponent was sitting over me with all four aces. I made the contract. I
recall when Thomas Dewey visited Ann Arbor. He walked through the
center of the quad. People were out sunbathing and simply looked up as
he went by. It did not cause much of a stir.
\section{Veterans Hospital Jan 1948--May 1948}
\textbf{Q.} How was your time spent after you left Ann Arbor in December of 1947
and until you enrolled in statistics at the University of North
Carolina in the fall of 1948?

\textbf{A.} You recall that I had stayed in Ann Arbor in the fall of 1947 to
have the spot on my lung followed. In January 1948, I entered a
Veterans Administration Hospital in Framingham, Massachusetts for
possible treatment. There I spent time in a facility for tuberculosis
patients, waiting ``recovery'' to start my graduate studies. The spot was
actually getting\break smaller, and the fact that it was changing caused
greater concern about my condition. Eventually I was released when it
was decided that I did not have or was clear of tuberculosis.

\textbf{Q.} How did you keep occupied in the hospital; did you play poker?

\textbf{A.} No, I took the time to study end games in chess. I~did not play
chess but enjoyed the study of end games. I treated end game play
puzzles as mathematics. I~did not read mathematics when at Framingham.
We were kept in bed and nourished. I was there about 2 months. Toward
the end I was allowed to leave on weekends---I~recall going to Boston
on a couple of weekends. I~did not have a car.

\section{UNC Aug 1948--Aug 1950}
\textbf{Q.} Did you enter the University of North Carolina in the fall of 1948
with an assistantship?

\begin{figure}[b]

\includegraphics{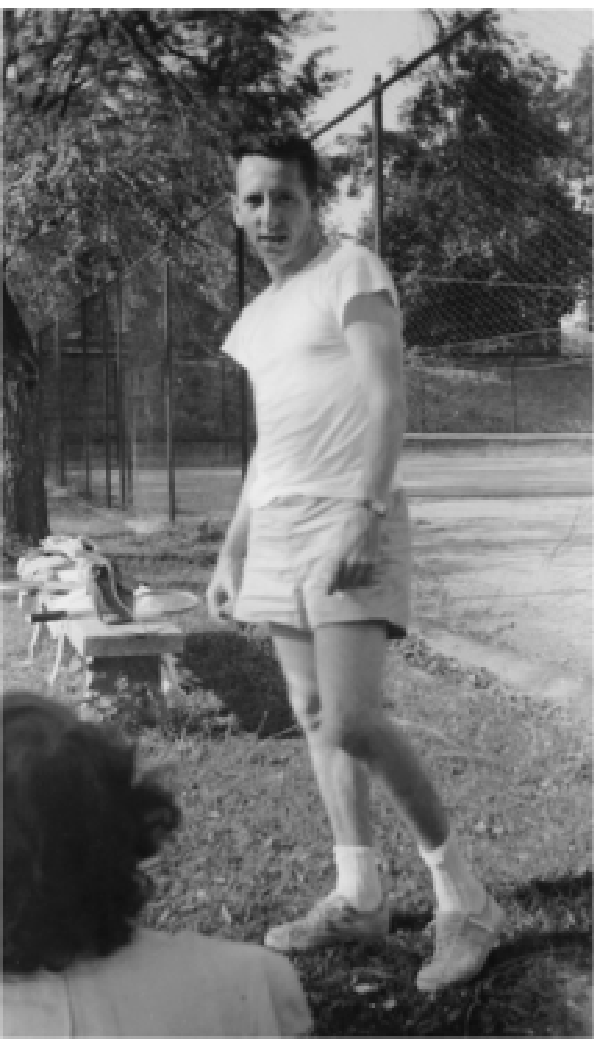}

  \caption{Jim at UNC, 1949.}
\end{figure}

\begin{figure*}

\includegraphics{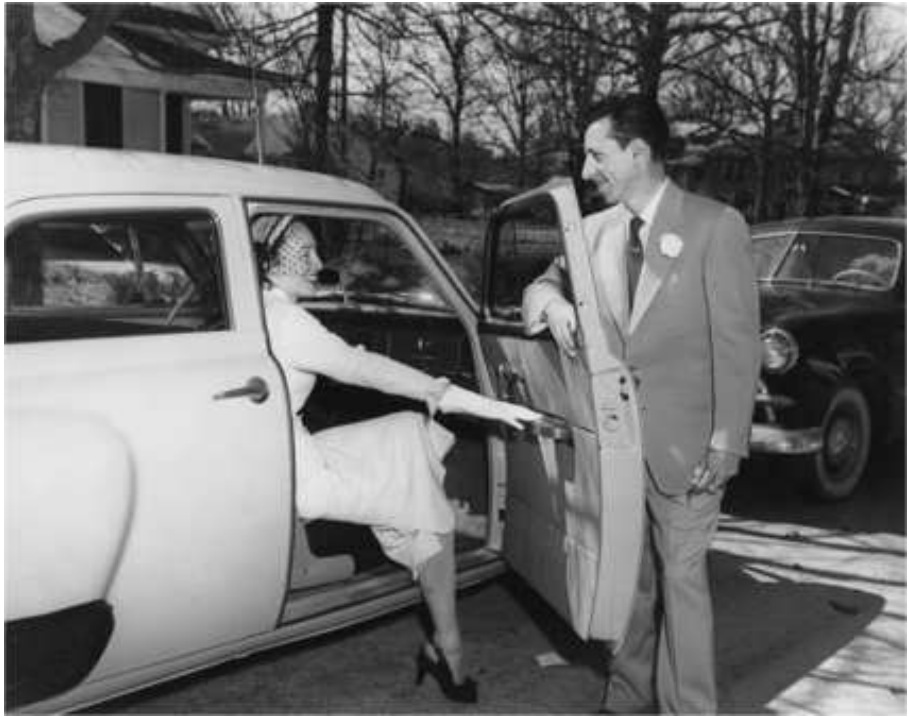}

  \caption{Bettie and Jim on wedding day, 1951.}
\end{figure*}

\textbf{A.} No. But I had support under the GI Bill and from my poker winnings.
I was soon given a research assistantship, more of a fellowship. They
had no need for teaching assistants.

\textbf{Q.} Who were your fellow students there at UNC?

\textbf{A.} Raj Bahadur, Sutton Monro, Ralph Bradley, Ingram Olkin,
Shrikande,\,\ldots.

\textbf{Q.} We understand that you met your future wife, Bettie Meade Creighton,
at UNC.

\textbf{A.} She was a graduate student working as a secretary to Hotelling,
having worked previously for the Dean of the Graduate School. Hotelling
was impressed with her considerable skills with dictation and
shorthand. We were married in 1951.

\textbf{Q.} What courses did you take during your two years at UNC?

\textbf{A.} The courses were small at UNC with ten or so students per class. I
recall taking a course in econometric models and least squares from
Harold Hotelling. He was more or less at the end of his career. I~did
learn why if you find five shoe stores in a mall, four of them will be
clustered close together. I took two courses from E.~J.~G. Pitman. He
had mathematics and Pitman ``efficiency.'' I took probability and
applications from Herbert Robbins my second year. Of course, he became
my thesis director. I had S. N. Roy for a course. Wassily Hoeffding
taught nonparametrics. He influenced me considerably and was a serious
reader of my thesis that came three years after I left UNC. I taught
Hoeffding how to drive while at UNC. Years later, I taught Pitman how
to drive while visiting Stanford. He had to keep being reminded that in
the United States we drive on the right-hand side of the road.

\textbf{Q.} Why did you leave before writing your dissertation?

\textbf{A.} I decided that I needed teaching experience and went to Catholic
University in Washington, DC. There I taught abstract algebra from
Birkhoff and Mac Lane, a course in differential equations and a reading
course in statistics.

\textbf{Q.} Your 1953 UNC thesis \cite{hannan1} contains several results, some in compound
decision theory for sets of statistical decision problems, some for
symmetrizations of product measures, some concerning averages of
empirical distributions from independent but not necessarily
identically distributed random variables. It is curious that your later
work in repeated games published in 1957 is directly relevant to
forecasting, specifically, selecting the best forecaster from a set of
forecasters based on the history of errors.

\textbf{A.} When I went to Robbins, I explained my interest in the challenges
posed from pooling across sequences of decision problems, perhaps,
having the C. C. Craig challenge in mind. Robbins quickly determined
that he could deal with sets of decision problems. His results on
compound decision theory were published in the \textit{Second Berkeley
Symposium} \cite{robbins51} and part of my thesis dealt with bounds and rates in
compounding general two-state problems.

\textbf{Q.} What topics did you study under Robbins?

\textbf{A.} Robbins was teaching probability and analysis, but he was interested
in certain applications of probability and probably ended up
concentrating on those and not going as deeply into general theory. He
had a free hand and taught whatever he wanted.

\textbf{Q.} At some point in the spring of 1950 you secured the position at
Catholic University. At that point did you have a thesis topic in mind?

\textbf{A.} Yes, more or less to extend Robbins' result on compound decision
theory with a general proof for the two-state problem. I believe that
Robbins published his compound decision problem in
the \textit{Second Berkeley Symposium}. The only inversion was that Robbins had
both the empirical Bayes formulation and the set compound formulation
in mind in 1950. He could only give one talk and thought that the
compound decision result would create a bigger splash. He gave the
empirical Bayes talk at the \textit{Third Berkeley Symposium} \cite
{robbins56}. (For an introduction to and review of the compound and
empirical Bayes decision problems, see Zhang~\cite{zhang}.)\looseness=1

\textbf{Q.} You mentioned talking to Robbins in your second year about
compounding in the sequence case. How did he react?

\textbf{A.} Robbins mentioned that, with the entire set of data available before
all decisions were made, one could estimate the empirical distribution
of the states. But that when only initial segments were available at
each stage, the problem is more difficult.

\textbf{Q.} In his \textit{Second Berkeley Symposium} paper, Robbins does give you credit
for some computations so you must have been actively involved with the
compound problem before leaving in the summer of 1950?

\textbf{A.} Yes, but the computations were wholly of interest to him because
they concerned the example he had of testing one specific normal versus
another and he needed the computations.

\textbf{Q.} You mentioned C. C. Craig at the University of Michigan challenging
the students to explain how the information coming across days from the
rattling of coins could be pooled to improve estimates. When you first
discussed pooling (compounding) with Robbins, did the discussion
trigger something in Robbins or was he already on the trail of compounding?

\textbf{A.} Well, he was on the trail of a method but claimed that he could not
do it. This was a strong statement that, of course, was wrong. He could
and did demonstrate it.

\textbf{Q.} For the set compound formulation not the sequence version?

\textbf{A.} Yes. I believe that Robbins did not push the sequence version of
compound decision theory until he had Esther Samuel as a graduate
student at Columbia. After she had come to the point of finishing her
thesis, Robbins suddenly remembered that I had done something on the
sequence problem. He became worried about her losing her thesis and
contacted me. She did not.

\textbf{Q.} Did Robbins get into the compound problem in his course?

\textbf{A.} Well, he was trying to talk about it.

\textbf{Q.} Your thesis dealt with the compounding across a set of two-state
(testing simple v. simple) decision problems.

\textbf{A.} Yes, and some other things.

\textbf{Q.} The results in symmetrizations of product measures?

\textbf{A.} Yes, that relates to the second chapter. However, I didn't put it
together as a final result with the strong properties that I sought.

\textbf{Q.} Did you ever publish that work on symmetrizations? We recall that
some of your students may have worked on this as well.

\textbf{A.} I did not publish it at the time, but I did get several of my Ph.D.
students onto some of the remaining research challenges.

\textbf{Q.} The symmetrizations result was a critical tool in showing the
asymptotic closeness of the simple and equivariant envelopes and in
attacking the Robbins' conjecture that eBayes procedures would solve
his compound problem.

\textbf{A.} Yes, that is true.

\textbf{Q.} At the time you left UNC after only two years of study, was your
thesis coming together?

\textbf{A.} No, I was struggling and not getting the results I thought I could
get. However, I already had accomplished the Kolmogorov--Smirov work
that was later to become Chapter 3 of my thesis.

\textbf{Q.} What were your classmates at UNC researching?

\textbf{A.} Bahadur was trying to interest Robbins and as soon as Robbins got a
glimpse that Raj was quite talented, he became interested. Raj had his
own problem, the ``problem of the greater mean.''

\textbf{Q.} What about the others?

\textbf{A.} I believe that Shrikande worked with R. C. Bose. Bradley may have
worked with Hotelling.\break Vohra was an early Ph.D. and I think that he
worked with Hoeffding. Hoeffding ended up with a couple of good
students the year after I left. Monro was there. He became interested
in the Robbins' inference problems. Ingram Olkin was a graduate student
during my first year.

\textbf{Q.} Was there a social life there involving the graduate students?

\textbf{A.} There was sort of a split. There was an Indian unit among the
graduate students including Bahadur, Shrikande and Chanti Vohra. The
natives were a motley crew including Bradley, Monroe and myself. There
was a split in the sense that the Indians had deeper training. Ralph
Bradley may have been an exception, but he was Canadian.

In Chapel Hill at that time, the Indians were a special case. Chapel
Hill was segregated then. Some got into private homes where the
landladies made sure that they were to be thought of as Indians and not
blacks. Bahadur resided in graduate housing, and it was the persons in
graduate housing that I got to know. I rented a room in a private home
that was near to the building that housed statistics.

\begin{figure}[b]

\includegraphics{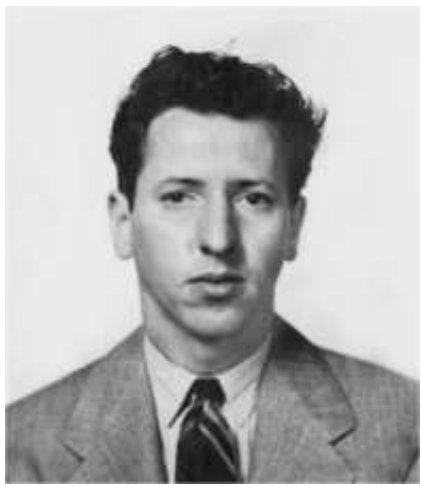}

  \caption{Jim's MSU Application Photo, 1953.}
\end{figure}

\begin{figure*}

\includegraphics{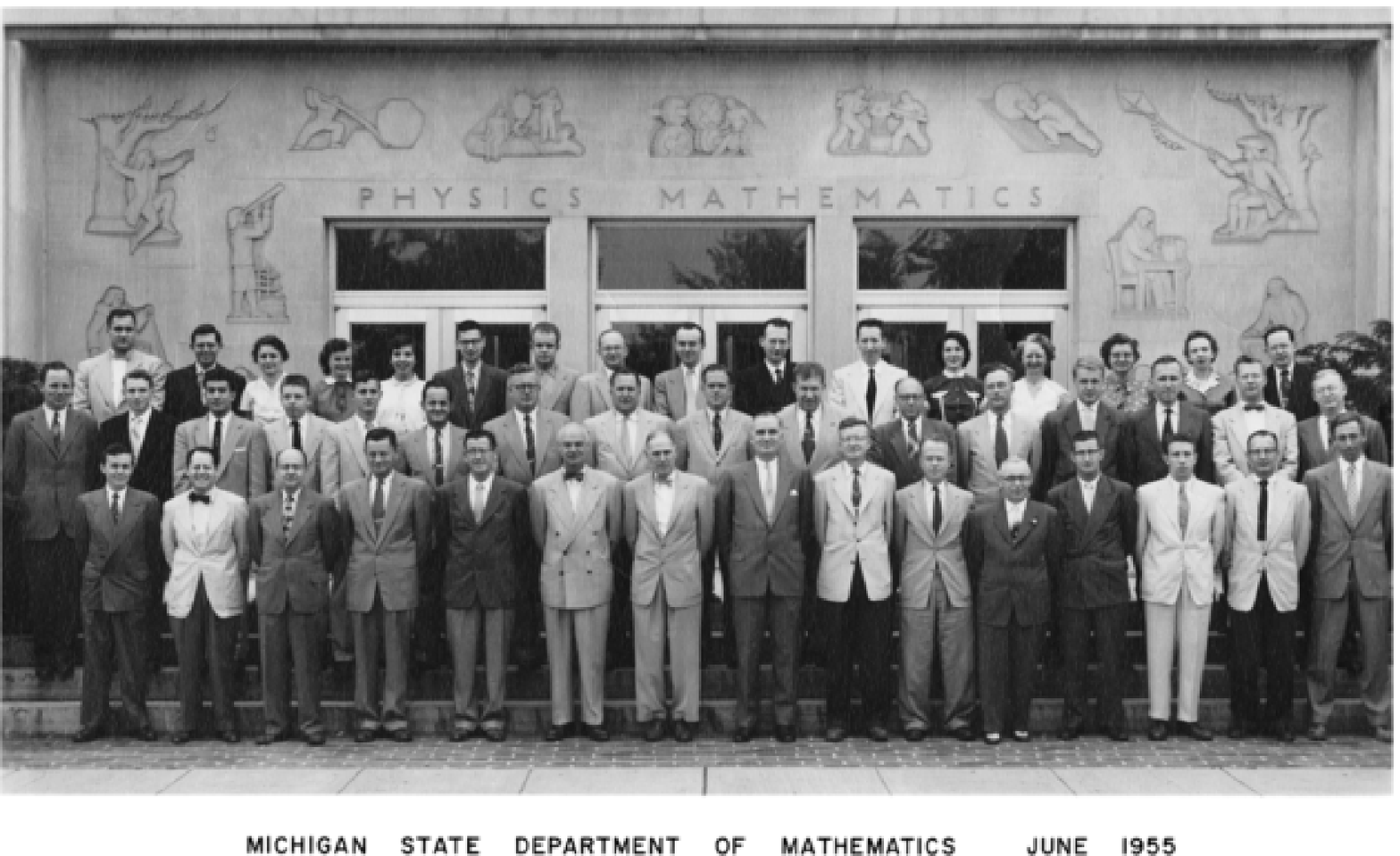}

  \caption{Michigan State University Department of Mathematics,
1955. Bottom row: 8 from right---Sutherland Frame, 4 from right---William Harkness,
 3 from right---Ingram Olkin. Second row: 6 from right---Leo Katz,
  5 from right---Ken Arnold. Top row: 6 from right---Jim
Hannan.}
\end{figure*}

At some point George Nicholson replaced Hotelling as chair. During the
war, Nicholson had tent mates on Guam that were statisticians in
operations research. When I left I was sharing an office with Raj
Bahadur and even his hours. I would stay up half the night (laugh).

\textbf{Q.} Were you smoking then?

\textbf{A.} Yes, and Raj was as well.

\textbf{Q.} Did you play tennis at that time.

\textbf{A.} Yes, with Raj and he was more adept at it than I.

\textbf{Q.} In Robbins' \textit{Second Berkeley Symposium} paper \cite{robbins51}, he gives
examples of and motivations for compounding. Did you leave UNC with the
idea that you would develop compound decision theory to cover general
component problems?

\textbf{A.} As you know, the compounding part of my thesis concerned two-state
problems with some of the work published with Robbins as co-author
\cite{hannan55}. I handed-off the m-state problem to John Van Ryzin,
one of my first graduate students, and with him published the paper on
improved rates for compounding the two-state problem~\cite{hannan65}.

\section{Michigan State University Sept~1953--Present}
\textbf{Q.} How did you happen to come to the Department of Mathematics at
Michigan State University in 1953?

\textbf{A.} Leo Katz was a visitor at UNC from Michigan State University and he
encouraged me to apply. Also Ingram Olkin was now at Michigan State
University, and he encouraged me to apply. Olkin was a Columbia M.S.
and took his Ph.D. from UNC.

\textbf{Q.} We have had a chance to read the letters of recommendations from
Robbins, Hotelling and Hoeffding in support of your application for an
asssitant professorship at Michigan State University. Were you the top
student in your cohort at UNC?

\textbf{A.} Remember that Bahadur was not in those classes. That is one
explanation for the strong letters. The students from India came with
strong backgrounds and were advanced to the point that they did not
take some of the classes that I took.

\textbf{Q.} Did you interview at places other than MSU?

\textbf{A.} In those days you did not go out and interview. Schools were hunting
for personnel rather than the other way around.

\textbf{Q.} It is curious that your MSU application form asked the applicant
about high school teaching experience. You admitted to none. When you
arrived what was the teaching load?

\textbf{A.} It was 12 hours/week, although some senior persons might have had
less. We were on the quarter system and the courses could be either 3
or 4 credits. The slide rule course was a 1 credit course. When I came,
Sutherland Frame was chair of the Department of Mathematics. I had used
Frame's pre-calculus textbook at St. Michael's.

\textbf{Q.} Who were the members of the statistics group when you arrived?

\textbf{A.} When I came, the statistics group included Leo Katz, Ken Arnold and
Ingram Olkin. Baten had an appointment in the Experiment Station as a
very applied statistician. It was rather shocking when he gave a talk
in which he described taking a random sample of trees and saying that
they came out ``real random.'' Oh, Charles Kraft came the next year. He
had left Berkeley because of the strike at Berkeley related to the
government's anti-American activity search. He did a thesis with LeCam.
Gopi Kallianpur came in 1956, and he came down with tuberculosis. I
recall visiting him in a Lansing hospital. Jim Stapleton and Martin Fox
came in the late 1950's. Herman Rubin and Esther Seiden came a little
later. Kraft and Olkin were only in the department a short time.

\textbf{Q.} So you were here when the Department of Statistics was created in 1955.

\textbf{A.} Yes. Frame served as chair of both departments that first year.
There was a dichotomy. It was recognized that the statistics faculty
could teach elementary math courses, while the math people could not or
were uninterested in teaching elementary statistics courses.

\textbf{Q.} One of the documented reasons for the creation of the Department of
Statistics was the fact the faculty in statistics would be expected to
consult and would, therefore, have a lesser teaching load. Did you have
consulting opportunities?

\textbf{A.} Yes, early on I did work somewhat collaboratively with Clifford
Hildreth of Economics. He was anxious to add more mathematics to his
research. This did not lead anywhere. He later became an Editor of the
\textit{Journal of the American Statistical Association}.

\textbf{Q.} Who were your early collaborators in basic research?

\textbf{A.} The first chapter of my thesis was published in 1955 in the \textit{Annals}
with Robbins as my co-author \cite{hannan55}. When I came to Michigan
State University in 1953, I began a collaboration with Jerry Gaddum,
who was in geometry. Blackwell had sent something of a mathematical
nature to Gaddum, and I helped him prove something. I then started
generalizing it so much that Gaddum was less happy, and he dropped out
of the research into repeated games. He was interested in linear
programming but not game theory. Gaddum is acknowledged in my 1957
paper in \textit{Contributions to the Theory of Games} \cite{hannan57}.

Early on I collaborated with Herman Rubin on a technical report titled
``Sigma-fields independent of sufficient sigma-fields.'' He okayed my
proof and had good knowledge of the literature that I did not have. I~did not work directly with others at MSU at that time. Over the years I
collaborated in research with a number of my graduate students and
others including Ronald Pyke, E.~J.~G. Pitman, R. F. Tate, Gordon
Simons, V\'{a}clav Fabian and R. V. Ramamoorthi.

\textbf{Q.} Perhaps now you are best known in the profession for your 1957 game
theory paper \cite{hannan57}. This paper is often referred to in
recognition of your having developed strategies for repeated play of a
game that have a property now called \textit{Hannan consistency}
\cite{hart}. At the same time, David Blackwell was establishing an
almost sure result using his approachability theory \cite
{blackwell56}. What was going on?

\textbf{A.} Of course, this research was related to some of my doctoral
research. The research under Robbins was on sets of statistical
decision problems. In the 1957 paper I was dealing with sequences of
games rather than sets of statistical decision problems. I did not
recognize approachability theory as applying. I let Blackwell know when
I completed the game theory work. He suggested that I publish my
results in the Princeton University Press series, which I did.
Blackwell's recognition of the use of his approachability theory to get
the envelope result came later, but he did get it into his publication
\cite{blackwell54}. Blackwell did alert me to the fact that
approachability theory could be used to prove results on convergence to
envelope risk. Someone told me, I have forgotten who, that Blackwell
had announced some results that either contained or were related to
mine and mentioned that I had priority. This was at a meeting somewhere.

\textbf{Q.} Have you ever used your game theory results in your personal decision-making?

\textbf{A.} I can't say that I have (laugh). I suspect that compounding is being
actively used. Economic theorists have been using it but without the
mathematics.

\textbf{Q.} Did you have leaves from MSU that took you away from East Lansing?

\textbf{A.} I went on leave to Stanford for the year 1956--1957. They had
openings for persons to work on contracts from the Office of Naval
Research. I did a few things related to those that didn't lead
anywhere. Herman Chernoff was sort of acting as an overseer of the
contract work. I recall that the contract work dealt with logistics of
keeping ships supplied. Too many persons were deciding what a ship
needed and my work was on creating discipline in inventory and stocking
for naval vessels. Some of the inventory problems were amenable to
compounding. Bellman was at the Rand Corporation and had already done
some related work but I did not make connections with him. Karlin had
been at Rand.

\textbf{Q.} Julia Robinson \cite{robinson} proved that in a two player game if
both players use the \textit{play against the past strategy}, then
average risk converges to the value of the game. This is called \textit
{fictitious play}. She credits Brown at the Rand Corporation for the
idea. Did you correspond with Robinson?

\textbf{A.} No. Her husband Raphael Robinson was on the mathematics faculty at
Berkeley. She did not get any recognition until the fictitious play
paper. Eventually they gave her an appointment. Neyman pushed her into
the National Academy of Sciences. His name carried some weight.

\textbf{Q.} Was there any reaction to your game theory work from the research
community? from the Rand Corporation?

\textbf{A.} Bowker and Karlin at Stanford were interested in the fact that I had
work related to Blackwell. People at Rand probably had limited command
of what Blackwell had done and an interest in other things that they
had better command of.

\textbf{Q.} Did you interact with others at Stanford?

\textbf{A.} Well, E. J. G. Pitman was there. He was at Chapel Hill my first year
and then at Stanford when I visited. I helped him find a car and taught
him how to drive on the ``wrong'' side of the road. He bought a car that
was ancient enough for him to feel comfortable with. His car had a
stick shift; he didn't want any additional complications. I worked with
him on a moments problem. We put out a technical report and did not
publish elsewhere. Birnbaum had a student who came up with similar
results at about the same time.

\textbf{Q.} Others?

\begin{figure*}

\includegraphics{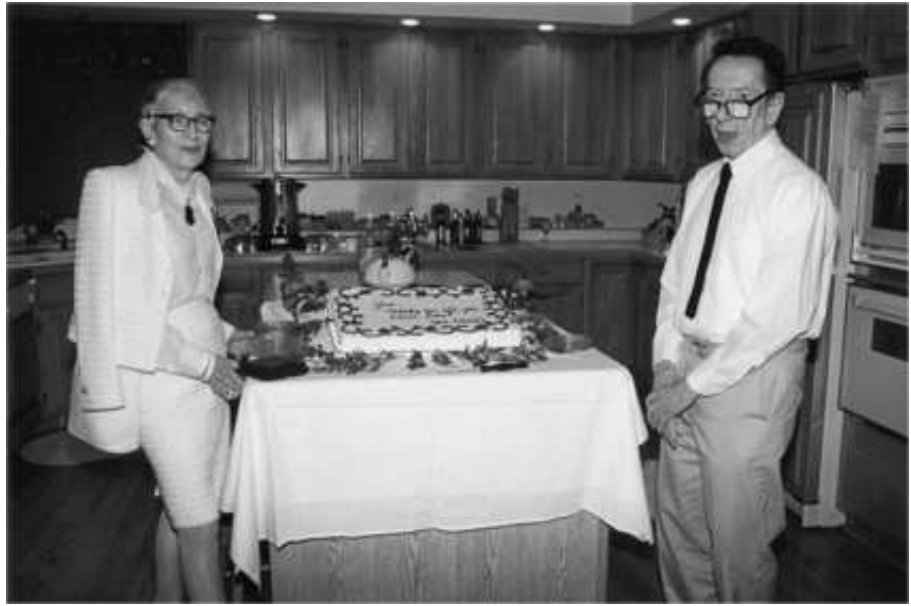}

  \caption{1998 celebration---Bettie and Jim.}
\end{figure*}

\textbf{A.} Samuel Karlin had two offices down the hall from me. We had rather
immediate contact in regard to his monotone likelihood ratio work. I
knew a few things and kept finding theorems where the proofs didn't
quite hold together. He may have stretched the Royden integration
theory a bit. I probably bothered him some. I recall him being anxious
to get to his bridge games that started at the lunch hour on Fridays
and sometimes lasted late into the afternoon. Vernon Johns was a
student of Robbins at Columbia and was then at Stanford as a regular
figure. He collaborated with a Donald Guthrie and also with Karlin and
Olkin on occasion. Ruppert Miller was a first-rate thinker and was
often the one to get things worked out.

\textbf{Q.} Did you meet Thomas Cover while at Stanford?

\textbf{A.} Yes, at a meeting and we spent two hours in a noisy room discussing problems.

\textbf{Q.} Did you teach at Stanford?

\textbf{A.} In addition to working on the project, I taught two quarter courses
plus a one quarter course in advanced inference using Lehmann's Notes.
I had almost all of the advance graduate students in my class including
William Pruitt and Donald Ylvisakar. Pruitt went to the University of
Minnesota and died rather young.

\textbf{Q.} Did you meet with David Blackwell while you were on the west coast?

\textbf{A.} I first met Blackwell when he invited Bettie and me to a lunch with
him and his family while I was visiting Stanford in 1956--1957. We
chewed up a lot of the family dinner. His wife was a very good cook and
she pushed lots of food on me. Anyway, when we met for lunch, it was a
long affair and he had committed to a seminar following it.

\textbf{Q.} Any other recollections of your time at Stanford?

\textbf{A.} When I gave a talk on the west coast, Charles Stein kept me on the
right track to finish my proof. I~refer to him in my game theory paper
for unpublished work on the prediction of a binary sequence. Stein was
at Stanford and the talk was probably at Berkeley. Stein's
inadmissibility result was presented at the Third Berkeley Symposium in
1955 \cite{stein}. Sometime later I sent Stein a note that I thought
that, in analysis of variance, the intraclass correlation would be a
case where compounding would apply. Subsequently, he published results
on this. The James--Stein \cite{james} and Efron--Morris \cite
{efron} connections to empirical Bayes came later.

\textbf{Q.} Was other research accomplished during your 1956--1957 leave to Stanford?

\textbf{A.} My research into maximum likelihood estimation of discrete
distributions was started at Stanford but was probably motivated
through discussions with Lucien LeCam. It was published in \cite
{hannan60} as a chapter in the Hotelling Volume.

\textbf{Q.} Was John Van Ryzin your first Ph.D. student?

\textbf{A.} No, earlier I had a third of William Harkness with Katz and Olkin. I
also had a half of Shashikala Sukatme. Her husband was in our
department. She started with Olkin, and I took over when he left. John
Van Ryzin left short of thesis, but he finished more quickly than I did
after I left UNC. That was another one of these things that does not
explain itself adequately. He had actually worked on conditions for
improved rates for the two-state problem, but I had done it for the
m-state. We agreed to split it so that we published the two-state
jointly, and he published the m-state (set version) solely. I felt that
the latter was more likely to be accepted for publication.

\textbf{Q.} We notice that he moved quickly to the sequence version of the
m-state. It came out in the \textit{Annals} as Van Ryzin \cite{vanryzin}, the
same year as did his set version m-state \cite{vanryzin66}. Didn't
another of your students leave short of thesis?

\textbf{A.} Yes, V. Susarla went to the University of Wisconsin, Milwaukee, in
1969 and finished his thesis in 1970. Van Ryzin was at Madison at the
time and ended his career as chair at Columbia. Susarla and Van Ryzin
became collaborators, and, tragically, they both lived short lives [Van
Ryzin (1935--1987) and Susarla (1943--1989)].

\textbf{Q.} In their memory you started the Van Ryzin--Susarla Fellowship for
study in the Department of Statistics and Probability at Michigan State
University.

Over the years you collaborated with some of your former graduate
students in resolving Robbins' (1951) conjecture that decision rules
that are Bayes versus diffuse priors in the empirical Bayes problem
will be asymptotic solutions to both his empirical Bayes and compound
decision problems. How did this research evolve?

\begin{figure*}

\includegraphics{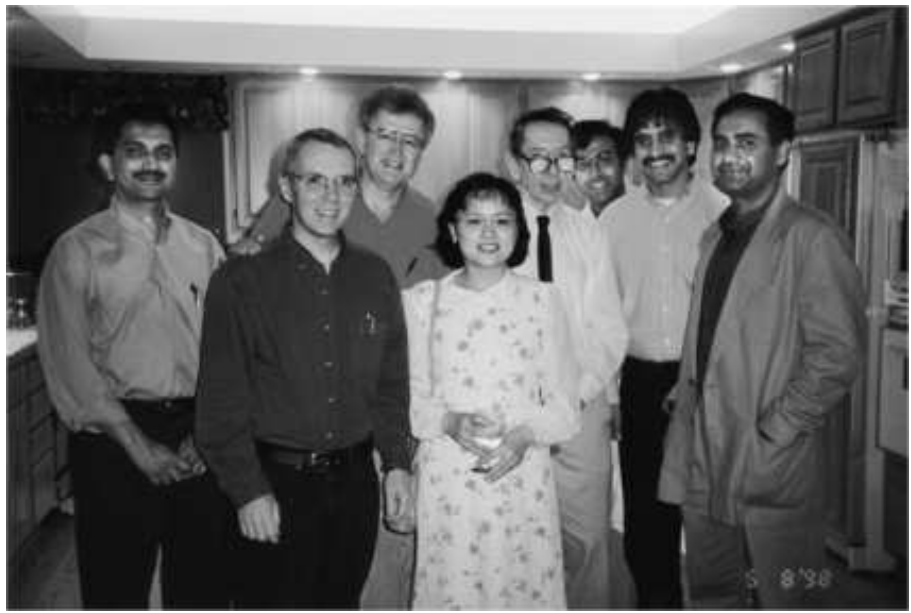}

  \caption{1998 with former students.
From left to right: J. Gogate, S. Vardeman, D. Gilliland, Y. H. Lai, J.
Hannan, S. Majumdar, M.~Mashayekhi, S. Datta.}
\end{figure*}

\textbf{A.} The research started in the early 1970s and appeared in technical
reports and theses. Published results with rates for finite state cases
came from collaborations with Gilliland \cite
{hannan85,hannan76,hannan69}, Huang \cite{hannan76,hannan72} and
Majumdar \cite{majumdar99}. In other more general contexts, rateless
related results were established by Datta \cite{datta91a,datta91},
Mashayekhi \cite{mostafa} and Majumdar \cite{majumdar2007}.

\textbf{Q.} You have recently reexamined the implications of results described
in your 1956 abstract \cite{hannan56}. How is this going?

\textbf{A.} In that abstract, I describe an extension of my repeated game
results to the case of repeated statistical decision problems. That
extension uses estimates for the empirical distribution of opponent's
past play and has implications for the less-than-full monitoring case
described in Rustichini \cite{aldo}. The estimation issues naturally
arise in compound decision problems. Vardeman \cite
{vardemana,vardeman} works this out for the k-extended envelope problem
\cite{hannan69}.

\textbf{Q.} You collaborated with V\'{a}clav Fabian in research and in writing
your 1985 book \textit{Introduction to Probability and Mathematical
Statistics} \cite{fabian}.

\begin{figure*}

\includegraphics{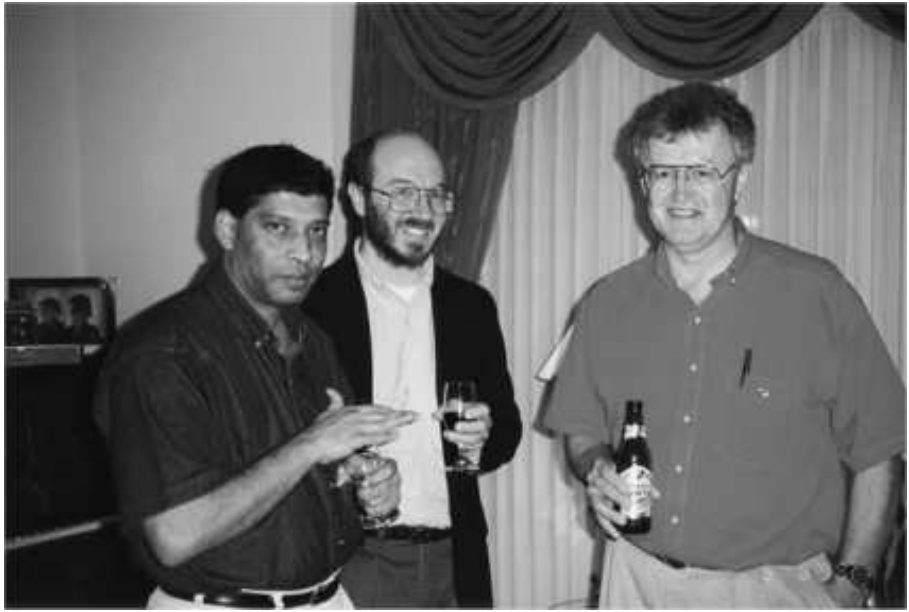}

  \caption{1998 colleagues---R. V. Ramamoorthi, Vince Melfi,
Dennis Gilliland.}
\end{figure*}

\textbf{A.} It was a very inefficient collaboration because we were never
working together on the same thing for the book. I anticipated
disagreement with Fabian's writing style but could not hold out, only
argue for my style.

\textbf{Q.} Bettie helped you have access to the Russian literature before it
was available in English translation. How did that come about?

\textbf{A.} Bettie was a graduate student in Art History at UNC but had no
education in modern languages. Here at MSU she became very interested
in French and then Russian. The latter was partially inspired by my
need for translations.

\textbf{Q.} Who were your teachers who most impressed and influenced you as a teacher?

\textbf{A.} I think first of Richard Bauer for group representations. He was a
magnificent lecturer. He would come in and do the review of the week in
5 minutes and it would be clear! He would then add to it. I was not
appreciating fully all that was involved, but it all fit together so
beautifully. I was naturally impressed with it all---it was such a
large difference from what I had had in the past. Andr\'{e} Gleyzal at
St. Michael's was slightly impressive, but he was holding back very
much from what he could do. His thesis came from Ohio State University
in mathematics.

\textbf{Q.} Concerning your own style, you are renowned for the concise notation
and connecting arrows in your board work.

\textbf{A.} I was not able to write fast on the chalkboard so I was
accommodating that fact through my style.

\textbf{Q.} You advised that one should not referee more papers than one submits
because refereeing should be taken seriously. You had a reputation as a
serious reader and did more than your share of work to improve papers
and theses. In regard to other service to the profession, how long were
you the book review editor of the \textit{Annals of Mathematical Statistics}?

\textbf{A.} I do not recall. (It was 1966--1972, seven full years.) I do recall
that Jack Keefer advised me to get competent reviewers and not to
heckle them.

\textbf{Q.} We have heard the Sir Ronald Fisher visited Michigan State
University in the late 1950s and gave a series of talks. What was that like?

\textbf{A.} He gave a series of lectures. He had a new book (recent at that
time) and wanted to talk about the book the way he thought about the
book and, Kraft, in particular, was more interested in mathematical
content. We had a conflict there. Kraft had definite theorems in mind,
and he could never get anything like that out of Fisher.

\textbf{Q.} So Kraft spoke up regularly during the lectures?

\textbf{A.} Well, he didn't badger him very much, but I knew what he had in
mind. He worked with LeCam and worked at that level. Fisher was still
back into thinking that everything he did was more important than
anything any one else did. (Laugh) I think that Fisher had to maybe
miss a few dates and then he came back. He threw away his notes and
proceeded to tell us what he thought of the kind of statistics we were
doing. He said the Annals should be bundled up and deposited into ``yon
river,'' the Red Cedar.

\textbf{Q.} What would have given him this inspiration to suddenly go on the offense?

\begin{figure*}

\includegraphics{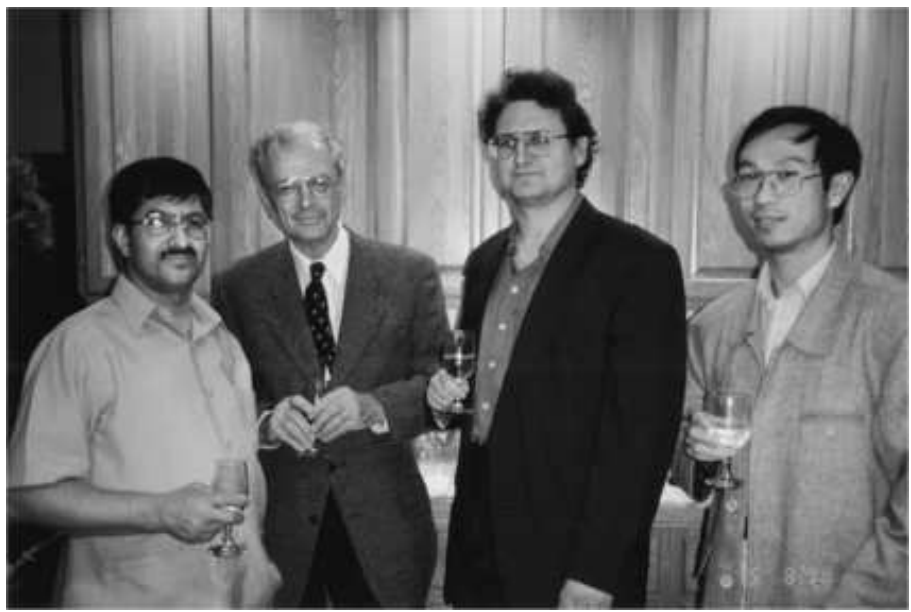}

  \caption{1998 colleagues---Hira Koul, Vaclav Fabian, Anton
Schick, Lijian Yang.}
\end{figure*}

\textbf{A.} Well, he was getting embarrassing questions. We were willing to talk
about his thing, but we wanted to talk about the mathematics of his
thing. Of course, eventually we noticed that we were not going to get
there and we shut up. It didn't get too nasty at that point, only after
he told us what he thought of us and what we could do with the \textit{Annals}.

\textbf{Q.} Did he mention Neyman in his talks?

\textbf{A.} I don't believe so. He had already fought that war.

\textbf{Q.} You are rather well known for your willingness to carefully read the
proofs of your professional colleagues. It seems that you often find
gaps, if not ways to improve proofs. Early in your career, you
discovered that Lemma 4.3.8 was false in the first printing of Samuel
Wilk's book \textit{Mathematical Statistics} \cite{wilks}. This
lemma was key to supporting the proofs of the asymptotic properties of
maximum likelihood estimators. How did you handle this discovery?

\textbf{A.} I took an interest in Wilks' book since I was looking for a strong
basis for asymptotics. I could not find it there. Wilks seemed to treat
analysis as algebra; I~know since that is where I started out. The
analysis was in Cram\'{e}r but he did not belabor the point. I still
have the correspondence. I wrote to Wilks in April 1962. I~realized
that the lemma was false and that Wilks was getting too much from too
little. He thanked me for the letter and advised that within six months
he would get back to me on the issues that I had raised. This he did.
In a letter in July 1962, Wilks writes, ``You are quite right. False as
stated. I am working on stronger conditions for the theorems that will
make them applicable to parametric estimation.''

\textbf{Q.} I understand that Wilks passed away before the second printing that
corrected the situation and acknowledged E.~J. Hannan for pointing out
the error.

\textbf{A.} Yes. Wilks' death was very sudden in March of 1964 at 57 years of
age. I believe that he died before having the opportunity to read the
proofs for the second printing. This version was completed by his
graduate students at Princeton, including David Freedman. There was a
very negative review of Wilks by Hoeffding \cite{hoeffding}.
Hoeffding never got emotional, but you can see that after a while in
this long review he had lost patience. Hoeffding was super conscious of
the fact that parametrization was not the be all and the end all in
statistics. I feel badly about my negative findings and the Hoeffding
review; these must have created a lot of stress.

\textbf{Q.} It is curious that your name causes much difficulty. Referring to
you as E.~J. Hannan is one of several instances. You are James Hanaan
in the bibliography of a Peyton Young's \cite{young} book. You are
James Hanna in Kolata's 2006 \textit{New York Times} article \cite{kolta}.

You seemed to be at your MSU office most every day of the week, even
after retirement in 2002. How many days of the year were you not at the office?

\begin{figure}[b]

\includegraphics{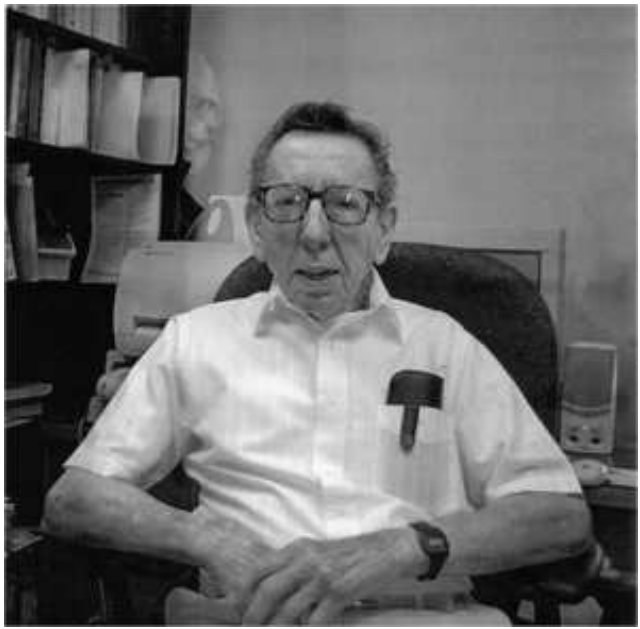}

  \caption{Jim, 2005.}
\end{figure}

\begin{figure}

\includegraphics{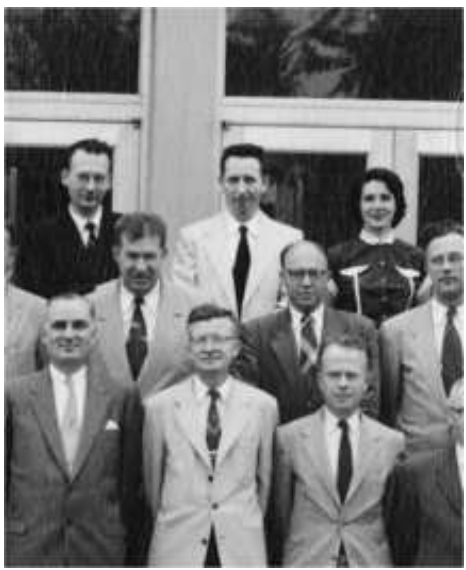}

  \caption{Michigan State University Department of Mathematics,
1955. Bottom row: On left---Sutherland Frame. Second row: 2 from left---Leo Katz,
 3 from left---Ken Arnold. Top row: 2 from left---Jim Hannan.}
\end{figure}

\textbf{A.} I did use to watch football on the weekends and on those days not
come to the office. I eventually quit watching football since, though
interesting, it was not useful. This was probably sometime in the 1960s.

\textbf{Q.} We recall a few times when you were impatient at faculty meetings.

\textbf{A.} In the early history we had Katz, Rubin and Kallianpur on the
faculty, persons with strong personalities. However, there are always
times when unanimity is not available or possible.

\textbf{Q.} Until recently you were heavily into exercise and kept a number of
us busy as workout partners. When did you get into this routine?

\textbf{A.} Probably after I quit smoking in the late 1960s. I~was really
hooked on cigarettes and coffee at the time. I quit coffee a month
ahead of quitting smoking. I~found it harder to quit coffee. To control
weight, I also quit my habit of having a large bowl of ice cream each evening.

\textbf{Q.} We observed that you would always attempt to surpass your workout
partners in regard to calorie burn and endurance. You were usually
successful regardless of the difference in ages. However, you failed to
top Marianne Huebner when she stood on her head for an extended period.

\textbf{A.} I was very impressed.

\textbf{Q.} You were not much for travel to meetings.

\textbf{A.} I did take two leaves to the west coast, and I usually went to Ann
Arbor for the joint colloquia. I did go to New York City for IMS
Meetings in 1971. That was about it for travel.

\textbf{Q.} Even after retirement you seem to be spending a lot of time in the
department. How do you spend time now?

\textbf{A.} I continue to study and work. I spent a lot of time carefully
studying the monograph on repeated games by Mertens, Sorin and Zamir
\cite{mertens}. Rustichini \cite{aldo} shows that some of my
results can be obtained via a result in this monograph. A much stronger
conclusion is possible with a weaker and easier version of the result
in MSZ. I am in the process of finalizing this. Often I start on a
paper and look up a reference. I find the reference interesting and
follow up on a reference to the reference. Thus, I may stray far away
from where I started.

\end{document}